\documentclass{icrc29}
\usepackage{graphicx,amssymb,amsmath,times}
\begin{document}
\title[Search for ...]{Search for multifractal features in cherenkov arrival time}
\author[A.Razdan] {A.Razdan$^a$\\ 
        (a) Nuclear Research Laboratory,
            Bhabha Atomic Research Centre,
            Mumbai 400 085, India \\
        }
\presenter{Presenter:A.Razdan (akrazdan@apsara.barc.ernet.in), \
ind-Razdan-A-abs2-og27-poster}

\maketitle

\begin{abstract}

Extensive air shower products are fractal in nature. Both simulated and experimental
Cherenkov images display multifractal properties. In this paper we explore the possibility
of searching multifractal features in cherenkov arrival times. 
\end{abstract}

\section {Motivation}

Extensive air shower (EAS) is a multifractal process because of multiplicative nature of
bremsstrahlung and pair production. It has been shown that EAS products like Cherenkov photons,
electron density distribution etc are multifractal in nature [2,3]. In this paper we search for multifractal
features in the temporal character of EAS.

\begin{figure}[h]
\begin{center}
\includegraphics*[width=0.8\textwidth,angle=270,clip]{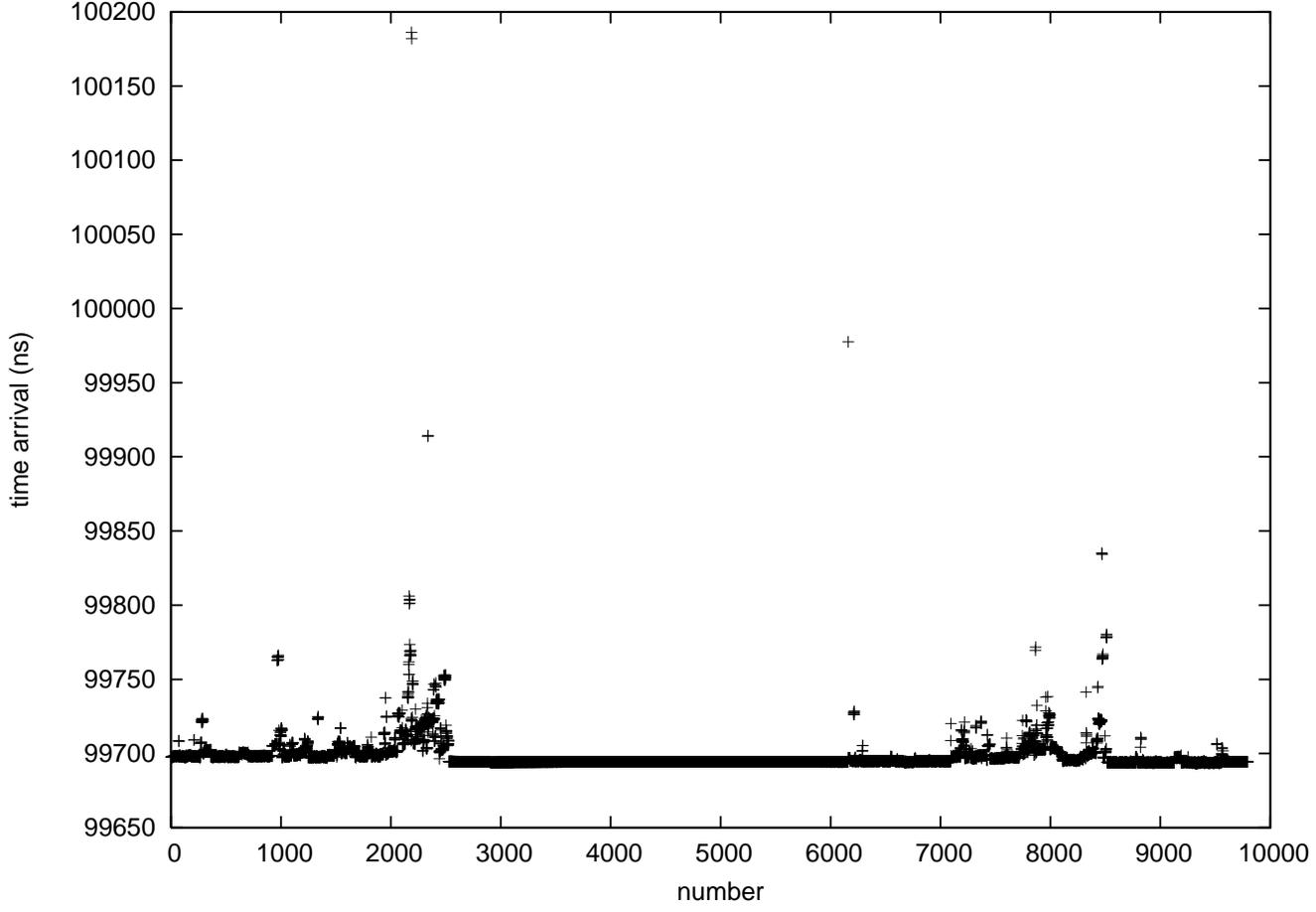}
\caption{\label {fig1} Cherenkov arrival time of a 2 TeV proton initiated shower.} 
\end{center}
\end{figure}

\section { Multifractal formalism}

Fractals are self-similar objects which look same on many different
scales of observations.
Fractals are defined in terms of Hausdroff-Bescovitch
dimensions. Fractal dimensions characterize the
geometric support of a structure
but can not provide any information about a possible distribution
or a probability that may be part of a given structure. This problem
has been solved by defining an infinite set of dimensions known as
generalized dimensions which are achieved by dividing the object under
study into pieces ,each piece is labelled by an index i=0,1,2....N. If
we associate a probability $p_i$ with each piece of size $l_i$ than
partition function is obtained which 
permits to define generalized dimension $D_q$
\begin{equation}
(q-1)D_q  =\tau(q)
\end{equation}
Here q is a parameter which can take all values between -$\infty$ to $\infty$.
$\tau(q)$ is obtained from scaling properties of the partition function.
This formalism is called as multi-fractal formalism which characterizes
both the geometry of a given structure and the probability measure
associated with it.
There are infinite set of other exponents
from which information can be obtained by constructing an equivalent
picture of the system in terms of scaling indices '$\alpha$' for the
probability measure defined on a support of fractal dimension f($\alpha$).
This is achieved by defining probability measure $p_i$ in terms of $\alpha$.
$\alpha$(q)) is the fractal dimension of the set.

\section {Approach followed in the present paper}

In this approach suggested by Chhabra et al [3]  whole experimental /simulation measure is covered with
boxes of size l and probability $P_i(l)$ is computed. From this probability construct a one parameter
family of normalized measure $\mu (q)$ 
\begin{equation}
\mu_i (q,l)= \frac{ [P_i(l)]^q }{\sum_{j}[P_j(l)]^q}
\end{equation}
 
The Hausdroff dimension of this measure is given as
\begin{equation}
f(q)= \lim_{l \rightarrow 0} \frac{ \sum_i (q,l) ln \mu_i (q,l)}{ ln l}
\end{equation}

The corresponding singularity strength $\alpha$ is given as
\begin{equation}
\alpha (q)= \lim_{l \rightarrow 0} \frac{ \sum_i (q,l) ln P_i (q,l)}{ ln l}
\end{equation}

As in the case of generalized dimensions, parameter q here also works as a microscope
to explore different regions of singular measure [3]. For q $>$1, $\mu(q)$ amplifies
the singular regions of the measure. For q $<$ , $\mu(q)$ accentuates the lesser regions of
singular measure. For q=1, the measure $\mu(1)$ replicates the original measure.
\begin{figure}[h]
\begin{center}
\includegraphics*[width=0.8\textwidth,angle=270,clip]{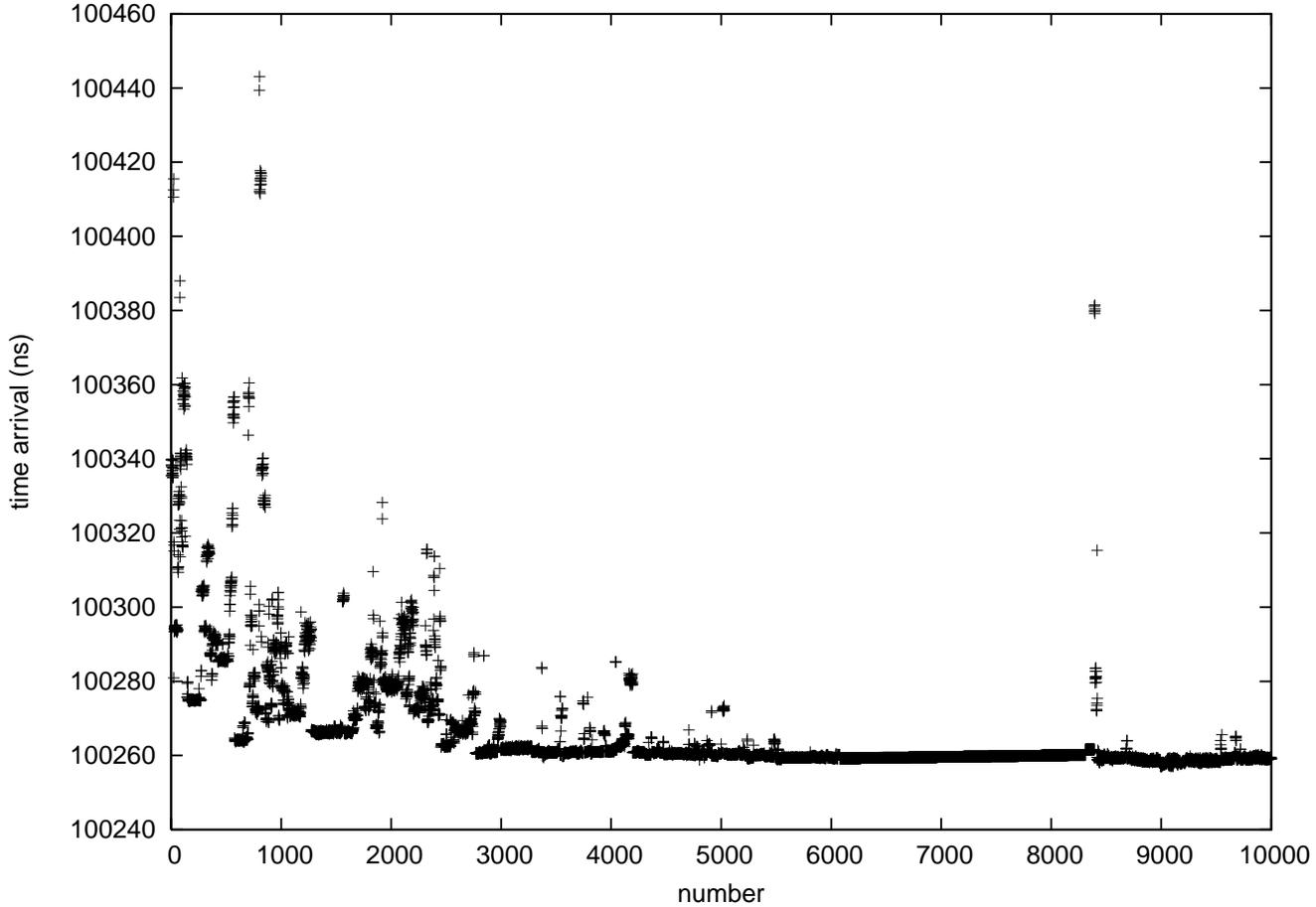}
\caption{\label {fig1} Cherenkov arrival time of 1 TeV $\gamma$-ray initiated shower.}
\end{center}
\end{figure}

\section {Simulation studies}

In the present studies simulations were carried out using CORSIKA (version 5.6211) along  with
EGS4, VENUS,GHEISHA codes for Cherenkov option. Simulated data is generated for
TACTIC [2] like configuration, each element of the size 4m X 4m. Simulated data corresponds to
Mt.Abu altitude ( 1300 m)  and appropriate  magnetic field. Cherenkov arrival time data corresponds to
wavelength band of 300-450 nm. 
\begin{figure}[h]
\begin{center}
\includegraphics*[width=0.8\textwidth,angle=270,clip]{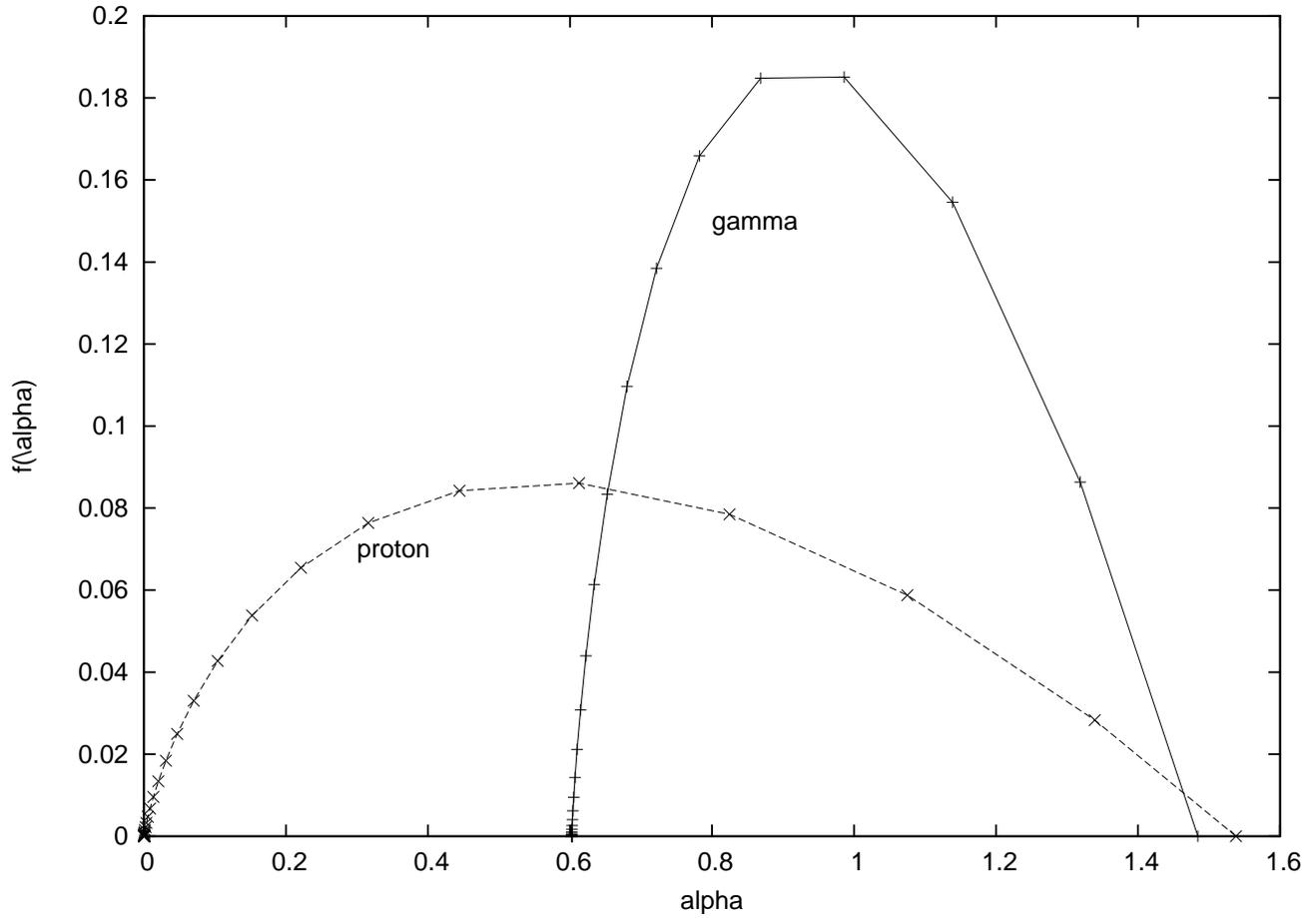}
\caption{\label {fig1} Multifractal spectrum of cherenkov arrival times}
\end{center}
\end{figure}
Cherenkov arrival time data for $\gamma$-rays and protons were generated for energies of 1 TeV and
2 TeV  respectively. It is observed that there are lot of fluctuations from shower to shower basis.
To do multifractal analysis we choose only that $\gamma$-ray and proton showers where number of photons
generated were roughly same. 
In cherenkov arrival times of $\gamma$-rays and protons  probability $P_i(l)$ was calculated for
various box sizes. By  calculating $\mu_i(q,l)$ in each box Hausdroff dimension and singularity 
strength was computed. It is clear from figure 3 that for both $\gamma$-ray and proton initiated showers
we obtain multifractal spectrum. f($\alpha$(q)) is the fractal dimension of the set. 

\section{ Conclusion}

In this paper it is shown that temporal character of EAS is  multifractal in nature. Earlier we had
shown that spatial character of EAS is multifractal [1,2]. So it can be concluded that EAS is a multifractal
process both in spatial and temporal manifestation.

\end{document}